\pdfoutput=1
\newcommand*{\ATLASLATEXPATH}{latex/}
\documentclass[cernpreprint, texlive=2016, UKenglish]{\ATLASLATEXPATH atlasdoc}

\usepackage{\ATLASLATEXPATH atlaspackage}
\usepackage{\ATLASLATEXPATH atlasbiblatex}

\usepackage{\ATLASLATEXPATH atlasphysics}

\graphicspath{{logos/}{figures/}}

\usepackage{pPbPhotonPaper-defs}

\AtlasTitle{Measurement of prompt photon production in $\sqrt{s_\mathrm{NN}} = 8.16$~\TeV{} $p$+Pb collisions with ATLAS}

\AtlasAbstract{%
The inclusive production rates of isolated, prompt photons in $p$+Pb
collisions at $\sqrt{s_\mathrm{NN}} = 8.16$~\TeV{} are studied with
the ATLAS detector at the Large Hadron Collider using a dataset with
an integrated luminosity of $165$~nb$^{-1}$ recorded in 2016. The
cross-section and nuclear modification factor $R_{p\mathrm{Pb}}$ are
measured as a function of photon transverse energy from $20$~\GeV\ to
$550$~\GeV{} and in three nucleon--nucleon centre-of-mass
pseudorapidity regions, $(-2.83,-2.02)$, $(-1.84,0.91)$, and
$(1.09,1.90)$. The cross-section and $R_{p\mathrm{Pb}}$ values are
compared with the results of a next-to-leading-order perturbative QCD
calculation, with and without nuclear parton distribution function
modifications, and with expectations based on a model of the energy
loss of partons prior to the hard scattering. The data disfavour a
large amount of energy loss and provide new constraints on the parton
densities in nuclei. }

\author{The ATLAS Collaboration}

\PreprintIdNumber{CERN-EP-2019-002}

\arXivId{1903.02209}

\AtlasJournalRef{Phys.\ Lett.\ B 796 (2019) 230}
\AtlasDOI{10.1016/j.physletb.2019.07.031}

\hypersetup{pdftitle={ATLAS document},pdfauthor={The ATLAS Collaboration}}

\begin{document}

\maketitle

\section{Introduction}
\label{sec:intro}

Measurements of particle and jet production rates at large transverse
energy are a fundamental method of characterising hard-scattering 
processes in all collision systems. In collisions
involving large nuclei, production rates are modified from those
measured in proton+proton ($pp$) collisions due to a combination of
initial- and final-state effects. The former arise from the dynamics of
partons in the nuclei prior to the hard-scattering process, while the
latter are attributed to the strong interaction of the emerging
partons with the hot nuclear medium formed in nucleus--nucleus
collisions. Modification due to the nuclear environment is quantified by 
the nuclear modification factor, $R_{\mathrm{AA}}$, defined as the ratio 
of the cross-section measured in A+A to that in $pp$ collisions, scaled by the 
expected geometric difference between the systems.

Measurements of prompt photon production rates offer a way to isolate
the initial-state effects because the final-state photons do not 
interact strongly. These initial-state effects include the 
degree to which parton densities are modified in a nuclear
environment~\cite{Salgado:2011wc,Arleo:2011gc,Helenius:2014qla},
as well as potential modification due to an energy loss arising through 
interactions of the partons traversing the nucleus prior to the hard
scattering~\cite{Vitev:2008vk,Kang:2015mta}. Constraints on such
initial-state effects are particularly important for characterising the
observed modifications of strongly interacting final states, such as jet
and hadron production~\cite{HION-2013-06,HION-2011-03},
since they are sensitive to effects from both initial- and final-state. 
Due to the significantly simpler underlying-event conditions in 
proton--nucleus collisions, measurements of photon rates can be 
performed with better control over systematic uncertainties than in 
nucleus--nucleus collisions, allowing a more precise constraint on these 
initial-state effects. 

Prompt photon production has been extensively measured in
$pp$ collisions at a variety of collision
energies~\cite{STDM-2010-08,STDM-2014-09,STDM-2016-08,CMS-QCD-10-037,CMS-QCD-10-035}
at the Large Hadron Collider (LHC). It was also measured in
lead--lead (Pb+Pb) collisions at a nucleon--nucleon centre-of-mass energy $\sqrt{s_\mathrm{NN}} =
2.76$~\TeV{}~\cite{HION-2012-02,Adam:2015lda} at the LHC, and in gold--gold
collisions at $\sqrt{s_\mathrm{NN}} =200$~\GeV{} at the Relativistic Heavy Ion Collider 
(RHIC)~\cite{Afanasiev:2012dg}, where the data from both colliders indicate that photon 
production rates are unaffected by the passage of the photons through the 
hot nuclear medium. At RHIC, photon production rates were measured 
in deuteron--gold collisions at $\sqrt{s_\mathrm{NN}} =200$~\GeV{}~\cite{Abelev:2009hx,Adare:2012vn}
and were found to be in good agreement with perturbative QCD (pQCD) calculations.
Additionally, jet production~\cite{HION-2013-08,CMS-HIN-14-001} and 
electroweak boson production~\cite{HION-2013-09,CMS-HIN-15-002,CMS-HIN-13-007}
were measured in $28$~nb$^{-1}$ of proton--lead ($p$+Pb) collision data at 
$\sqrt{s_\mathrm{NN}} = 5.02$~\TeV{} recorded at the LHC; the former is a strongly 
interacting final state, while the latter is not. All
measurements provided some constraints on initial-state effects.

The data used in this measurement were collected with the ATLAS detector
during the $p$+Pb collision running period in 2016, and correspond to
an integrated luminosity of $165$~nb$^{-1}$, approximately six times larger than 
the measurements made in the previous 5.02 TeV data. The proton and lead beams
had an energy of $6.5$~\TeV{} and $2.51$~\TeV{} per nucleon respectively, 
resulting in a nucleon--nucleon centre-of-mass collision energy 
of $8.16$~\TeV{} and a rapidity boost of this frame by $\pm0.465$ units 
relative to the ATLAS laboratory frame, depending on the direction of the 
Pb beam.\footnote{ATLAS uses a
  right-handed coordinate system with its origin at the nominal
  interaction point (IP) in the centre of the detector and the
  $z$-axis along the beam pipe. The $x$-axis points from the IP to the
  centre of the LHC ring, and the $y$-axis points upward. Cylindrical
  coordinates $(r,\phi)$ are used in the transverse plane, $\phi$
  being the azimuthal angle around the $z$-axis. The pseudorapidity is
  defined in terms of the polar angle $\theta$ as
  $\eta=-\ln\tan(\theta/2)$ and the rapidity of the components of the 
  beam, $y$, are defined in terms of their energy, $E$, and longitudinal 
  momentum, $p_z$, as $y=0.5\ln\frac{E+p_z}{E-p_z}$.} 
By convention, the results are reported as a function of photon 
pseudorapidity in the nucleon--nucleon collision frame,
$\eta^{*}$, with positive $\eta^{*}$ corresponding to the proton beam 
direction, and negative $\eta^{*}$ corresponding to the Pb beam 
direction.

At leading order, the process $p+\mathrm{Pb} \to \gamma + X$ has
contributions from direct processes, in which the photon is produced
in the hard interaction, and from fragmentation processes, in which it is
produced in the parton shower. Beyond leading order the direct and 
fragmentation components are not separable and only their sum is a
physical observable.

To reduce contamination from the dominant background of photons
mainly from light-meson decays in jets, the measurements presented 
here require the photons to be isolated from nearby particles.
This requirement also acts to reduce the relative contribution of 
fragmentation photons in the measurement, and thus, the same fiducial
requirement must be imposed on theoretical models when comparing with
the data. Specifically, as in previous ATLAS measurements~\cite{STDM-2014-09,STDM-2016-08}, 
the sum of energy transverse to the beam axis within a cone of 
$\Delta{R} \equiv \sqrt{ (\Delta\eta)^2 + (\Delta\phi)^2 } = 0.4$ 
around the photon, $E_\mathrm{T}^\mathrm{iso}$, is required to be smaller than 
$4.8 + 4.2\times10^{-3} E_\mathrm{T}^\gamma$~[\GeV{}], where 
$E_\mathrm{T}^\gamma$ is the transverse energy of the photon. At 
particle level, $E_\mathrm{T}^\mathrm{iso}$ is calculated as the sum of 
transverse energy of all particles with a decay length above 10 mm, 
excluding muons and neutrinos. This sum is corrected for the ambient 
contribution from underlying-event particles, consistent with the 
previous measurements~\cite{STDM-2014-09,STDM-2016-08}. 

This letter reports a measurement of the cross-section for prompt,
isolated photons in $p$+Pb collisions at $\sqrt{s_\mathrm{NN}} =
8.16$~\TeV{}. Photons are measured with $E_\mathrm{T}^{\gamma} >
20$~\GeV, the isolation requirement detailed above, and in three
nucleon--nucleon centre-of-mass pseudorapidity ($\eta^*$) regions,
$-2.83 < \eta^* < -2.02$, $-1.84 < \eta^* < 0.91$, and $1.09 < \eta^*
< 1.90$. In addition to the cross-section, the data are compared to 
a \pp reference cross-section derived from a previous measurement of 
prompt photon production in \pp collisions at $\sqrt{s} = 8$~\TeV{} that used the 
identical isolation condition ~\cite{STDM-2014-09}. The nuclear 
modification factor $R_{p\mathrm{Pb}}$ is derived in each pseudorapidity 
region, using an extrapolation for the different collision energy and 
center-of-mass pseudorapidity selection, and is reported in the region 
$E_\mathrm{T}^{\gamma} > 25$~\GeV where reference data is available.  
Furthermore, the ratio of
$R_{p\mathrm{Pb}}$ in the forward region to that in the backward
region is presented. The measurements are compared with
next-to-leading-order (NLO) pQCD predictions from
\textsc{Jetphox}~\cite{Aurenche:2006vj} using parton distribution
functions (PDF) extracted from global analyses that include nuclear
modification effects
analyses~\cite{Eskola:2016oht,Kovarik:2015cma}. Additionally, the data
are compared with predictions from a model including initial-state
energy loss~\cite{Vitev:2008vk,Kang:2015mta,Chien:2015vja}.

\section{Experimental set-up}
\label{sec:setup}

The ATLAS detector~\cite{PERF-2007-01} is a multipurpose detector
with a forward--backward symmetric cylindrical geometry. For this
measurement, its relevant components include an inner tracking
detector surrounded by a thin superconducting solenoid, and electromagnetic 
and hadronic calorimeters. The inner-detector system is immersed in a $2$~T 
axial magnetic field and provides charged-particle tracking in the pseudorapidity range
$\left|\eta^{\mathrm{lab}}\right| < 2.5$ in the laboratory frame. In order of 
closest to furthest from the beam pipe, it consists of a high-granularity 
silicon pixel detector, a silicon microstrip tracker, and a transition radiation tracker. 
Additionally, the new insertable B-layer~\cite{ATLAS-TDR-19} has been 
operating as the innermost layer of the tracking system since 2015.
The calorimeter system covers the range $\left|\eta^{\mathrm{lab}}\right| < 4.9$. In the
region $\left|\eta^{\mathrm{lab}}\right| < 3.2$, electromagnetic calorimetry is
provided by barrel and endcap high-granularity lead/liquid-argon (LAr)
sampling calorimeters. An additional thin LAr presampler covers
$\left|\eta^{\mathrm{lab}}\right| < 1.8$ to correct for energy loss in material 
before the calorimeters. The LAr calorimeters are divided 
into three layers in radial depth. Hadronic calorimetry is provided by a 
steel/scintillator-tile calorimeter, segmented into three barrel structures within
$\left|\eta^{\mathrm{lab}}\right| < 1.7$, and two copper/LAr hadronic endcap
calorimeters, which cover the region $1.5 < \left|\eta^{\mathrm{lab}}\right| <
3.2$. Finally, the forward calorimeter covers $3.2 <
\left|\eta^{\mathrm{lab}}\right| < 4.9$ and is divided into three 
compartments. The first compartment is a copper/LAr electromagnetic
calorimeter, while the remaining two tungsten/LAr calorimeter 
compartments collect the hadronic energy.

During data-taking, events were initially selected 
using a level-1 trigger, implemented in custom electronics, based on energy 
deposition in the electromagnetic calorimeter. The high-level trigger 
\cite{TRIG-2016-01} was then used to select events consistent with a
high-$E_\mathrm{T}^{\gamma}$ photon candidate. The high level trigger was configured with five online
$E_\mathrm{T}^{\gamma}$ thresholds from $15$~\GeV\ to $35$~\GeV. 
Each trigger is used for an exclusive region of the $E_\mathrm{T}^{\gamma}$
spectrum, starting 5~\GeV{} above the trigger threshold because there the trigger is 
fully efficient. The highest-threshold trigger is used in the measurement 
over the whole $E_\mathrm{T}^{\gamma}$ range above 40~\GeV{} and is unprescaled. 
The lower-threshold, prescaled, triggers are used to perform the measurement for
$E_\mathrm{T}^{\gamma}$ in the range of $20-40$~\GeV.

Data-taking was divided into two periods with different configurations of the 
LHC beams. In the first period, the lead ions circulated in beam 1 (clockwise) and
protons circulated in beam 2, while in the second period the beams were
reversed. These periods corresponded to integrated luminosities of $57$~nb$^{-1}$ 
and $108$~nb$^{-1}$ respectively. 

\section{Photon reconstruction and identification}
\label{sec:reco}

Photons are reconstructed following a procedure used extensively in
previous ATLAS measurements~\cite{STDM-2016-08}, of which only
the main features are summarised here. 

Photon candidates are reconstructed from clusters of energy deposited
in the electromagnetic calorimeter in three regions corresponding to
the laboratory-frame ($\eta^{\mathrm{lab}}$) positions of the barrel
and forward and backward endcaps $\left|\eta^{\mathrm{lab}}\right| <
2.37$.  The transition region between the barrel and endcap
calorimeters, $1.37 < \left|\eta^{\mathrm{lab}}\right| < 1.56$, is
excluded due to its higher level of inactive material.  The
measurement of the photon energy is based on the energy collected in
calorimeter cells in an area of size $\Delta\eta\times\Delta\phi =
0.075\times0.175$ in the barrel and $\Delta\eta\times\Delta\phi =
0.125\times0.125$ in the endcaps. It is corrected via a dedicated
energy calibration~\cite{Aaboud:2018ugz} which accounts for losses in
the material before the calorimeter, both lateral and longitudinal
leakage, and for variation of the sampling-fraction with energy and
shower depth.

The photons are identified using the tight calorimeter shower shape 
requirements described in Ref.~\cite{PERF-2017-02}. The tight requirements 
select clusters which are compatible with originating from a single photon 
impacting the calorimeter. The information used includes that from the 
hadronic calorimeter, the lateral shower shape in the second layer of 
the electromagnetic calorimeter, and the detailed shower shape in the 
finely segmented first layer. 

The isolation transverse energy, $E_\mathrm{T}^\mathrm{iso}$, is computed 
from the sum of $E_\mathrm{T}$ values in topological clusters of calorimeter
cells~\cite{PERF-2014-07} inside a cone of size $\Delta R=0.4$ centred on
the photon. This cone size is chosen to be compatible with a previous
measurement of photon production in $pp$ collisions at $\sqrt{s} =
8$~\TeV{}~\cite{STDM-2014-09}, which is used to construct the reference
spectrum for the $R_{p\mathrm{Pb}}$ measurement. This estimate 
excludes an area of $\Delta\eta\times\Delta\phi = 0.125\times0.175$ centred 
on the photon, and is corrected for the expected leakage of the photon 
energy from this region into the isolation cone.

\section{Simulated event samples}
\label{sec:MC}


Samples of Monte Carlo (MC) simulated events were generated to study 
the detector performance for signal photons. Proton--proton generators were 
used as the source of events containing photons. To include the effects of the $p$+Pb 
underlying-event environment, these simulated $pp$ events were combined with $p$+Pb 
events from data before reconstruction. In this way, the simulated events contain the 
effects of the $p$+Pb underlying-event identical to those observed in data. 

The \PYTHIA 8.186~\cite{Sjostrand:2007gs} generator was used to generate
the nominal set of MC events, with the NNPDF23LO parton distribution function 
(PDF) set~\cite{Ball:2012cx} and a set of generator 
parameters tuned to reproduce minimum-bias $pp$ events with 
the same collision energy as that in the $p$+Pb data (``A14'' tune)~\cite{ATL-PHYS-PUB-2014-021}. 
A centre-of-mass boost was applied to the generated events to bring them into the
same laboratory frame as the data. The generator 
simulates the direct photon contribution and, through final-state QED 
radiation in $2\to2$ QCD processes, also includes the fragmentation photon
contributions; these components are defined to be signal photons. 
Events were generated in six exclusive $E_\mathrm{T}^{\gamma}$ 
ranges from 17~\GeV\ to 500~\GeV. 

An additional MC sample was used to assess the sensitivity of the 
measurement to this choice of generator. The \SHERPA 2.2.4~\cite{Gleisberg:2008ta} 
event generator produces fragmentation photons in a different way from
\PYTHIA and was thus chosen for the comparison. The NNPDF3.0NNLO PDF set~\cite{Ball:2014uwa} 
was used, and the events were generated in the same kinematic regions as the \PYTHIA events.  
These events were generated with leading-order matrix 
elements for photon-plus-jet final states with up to three additional partons,
which were merged with the \SHERPA parton shower. The \SHERPA
sample produced results consistent with \PYTHIA, and, thus, no correction
or uncertainty is applied. 

The \PYTHIA and \SHERPA $pp$ events were passed through a full
\textsc{Geant}4 simulation of the ATLAS
detector~\cite{Agostinelli:2002hh,SOFT-2010-01}.
To model the underlying event effects, each simulated event was combined with a 
minimum-bias $p$+Pb data event and the two were reconstructed together as a 
single event, using the same algorithms as used for the data. These events were
split between the two beam configurations in a proportion matched to that
in data-taking. The underlying event activity levels, as characterized by the sum 
of the transverse energy in the outgoing-Pb-beam side of the forward calorimeter
(3.1 < $|\eta^{\mathrm{lab}}|$ < 4.9), are different in the photon-containing data 
events from the minimum-bias data events used in the simulation. Thus, the simulated 
events were weighted on a per-event basis to match the underlying event activity 
distribution in data. Furthermore, the photon shower shapes and identification 
efficiency in simulation were adjusted for small differences previously observed 
between these quantities in data and in \textsc{Geant}4 simulation~\cite{PERF-2017-02}.

\section{Data analysis}
\label{sec:analysis}

The differential cross-section is calculated for each $E_\mathrm{T}^{\gamma}$ and
$\eta^*$ bin as

\begin{equation*}
  \frac{ \mathrm{d}\sigma}{\mathrm{d}E_\mathrm{T}^{\gamma}} = \frac{1}{L_\mathrm{int}} \frac{1}{\Delta E_\mathrm{T}^{\gamma} } \frac{N_\mathrm{sig} P_\mathrm{sig} }{ \epsilon^\mathrm{sel} \epsilon^\mathrm{trig} } C_{\mathrm{MC}},
\label{eq:xsec}
\end{equation*}

where $L_\mathrm{int}$ is the integrated luminosity, 
$\Delta E_\mathrm{T}^{\gamma}$ is the width of the 
$E_\mathrm{T}^{\gamma}$ bin, $N_\mathrm{sig}$ is the
yield of photon candidates passing identification and isolation
requirements, $P_\mathrm{sig}$ is the purity of the signal selection, 
$\epsilon^\mathrm{sel}$ is the combined reconstruction,
identification and isolation efficiency for signal photons,
$\epsilon^\mathrm{trig}$ is the trigger efficiency, and $C_{\mathrm{MC}}$ is a
MC derived bin-by-bin correction for the change in the $E_\mathrm{T}^{\gamma}$ 
spectrum from photons migrating between bins in the spectrum due to the width 
in the energy response. $C_{\mathrm{MC}}$ is determined after all selection 
criteria at both reconstruction and particle levels are imposed.

Trigger efficiencies $\epsilon^\mathrm{trig}$ are studied using events
selected with minimum-bias triggers, level-1 triggers without
additional requirements, and photon high-level triggers without
identification requirements. They are greater than 99.5\% for 
all triggers~\cite{TRIG-2016-01}. In this analysis they are
taken as $\epsilon^\mathrm{trig} = 1$, and any uncertainty is neglected as 
being sub-dominant to other uncertainties.

The purity $P_\mathrm{sig}$ is determined via a double-sideband
procedure used extensively in previous measurements of cross-sections
for processes with a photon in the final
state~\cite{STDM-2014-09,STDM-2016-08,STDM-2012-16,STDM-2015-11}
and summarised here. In
the procedure, four regions are defined which categorise
photon candidates along two axes: (1)  isolation,
corresponding to an isolated and an inverted ``non-isolated'' selection;
(2) identification, corresponding to photons that pass the tight identification 
requirements described in Ref.~\cite{PERF-2017-02}, and those that pass the
{\em loose} requirements of Ref.~\cite{PERF-2017-02} but fail certain components
of the tight requirements, designed to mostly select background. 
The majority of signal photons 
are in the tight, isolated region, defined to be the signal region, while the other
three regions are dominated by the background. Photon candidates that
comprise the background are assumed to be distributed in a way that is
uncorrelated along the two axes. The yields in the three non-signal
sidebands are used to estimate the yield of background in the signal region
and is combined with the yield in the signal region to extract the purity.
The procedure also accounts for the small fraction of signal
photons which are reconstructed in the non-signal sidebands; these
quantities, known as leakage fractions, are determined from the simulation
samples described in Section~\ref{sec:MC}. The purity is typically 
45\% at $E_\mathrm{T}^{\gamma} = 20$~\GeV, rises to 80\% at 
$E_\mathrm{T}^{\gamma} = 100$~\GeV{} and reaches 99\% at 
$E_\mathrm{T}^{\gamma} = 300$~\GeV.

\begin{figure}[t]
\begin{center}
\includegraphics[width=\textwidth]{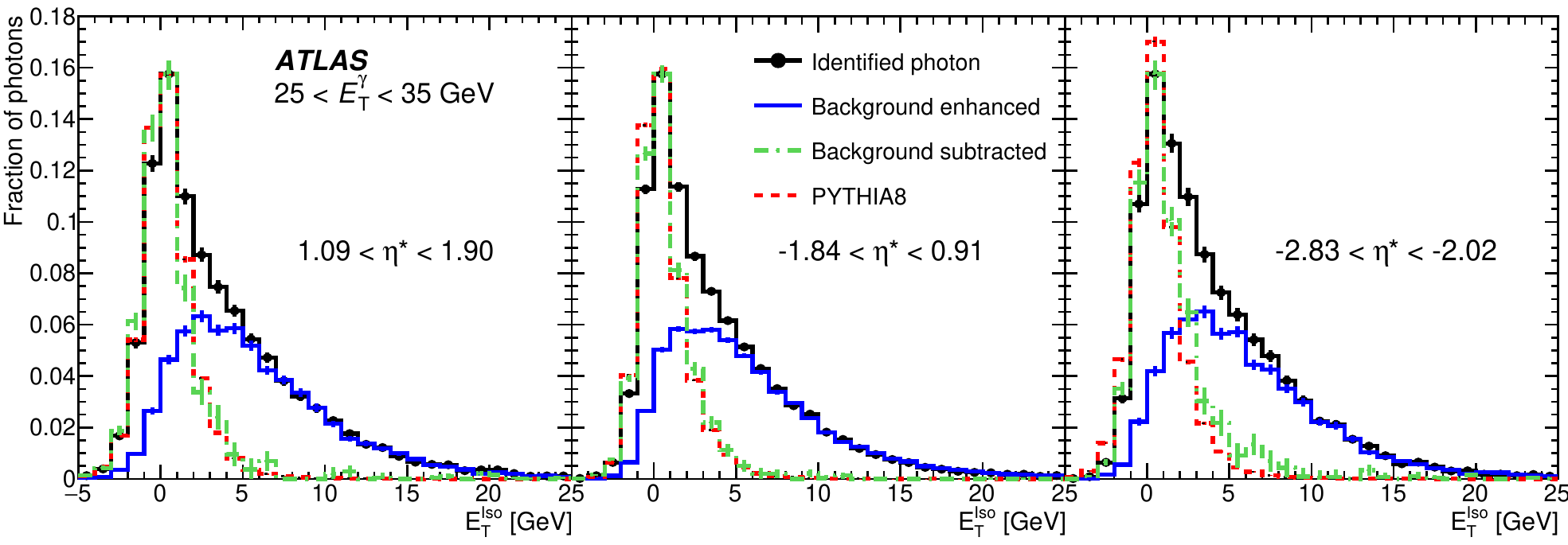}
\includegraphics[width=\textwidth]{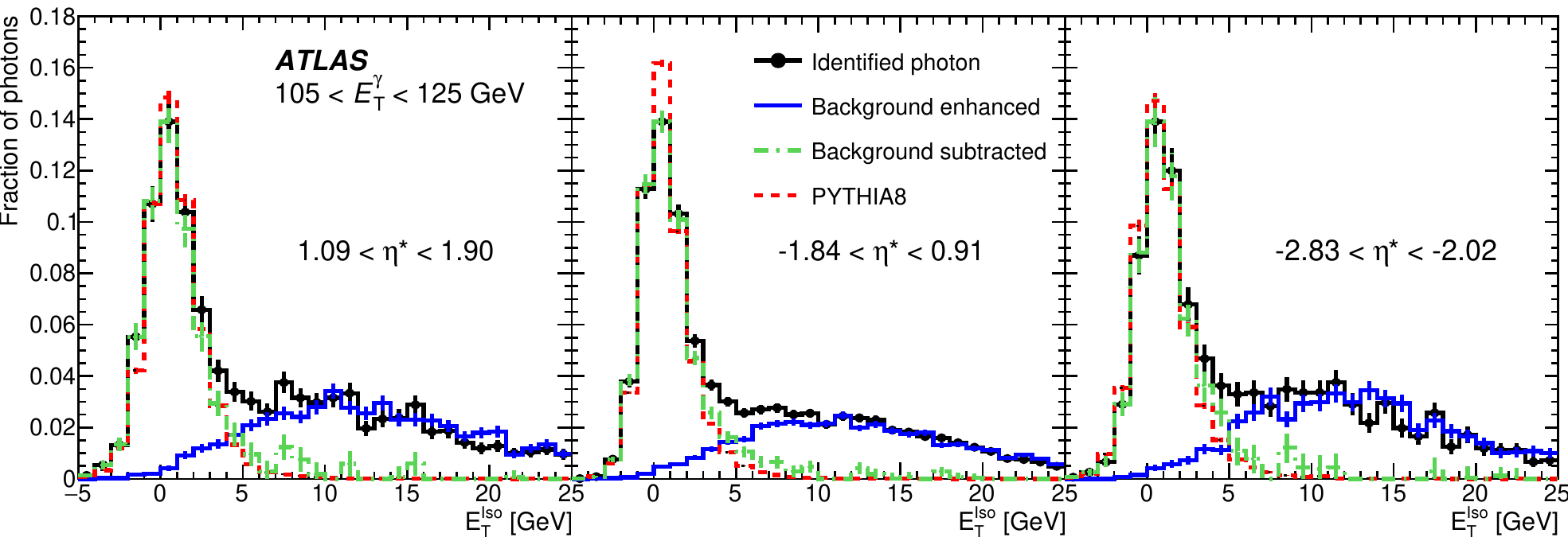}
\caption{Distributions of detector-level photon isolation transverse energy
    $E_\mathrm{T}^\mathrm{iso}$ for identified photons in data
    (black points), background photons scaled to match the data at
    large $E_\mathrm{T}^\mathrm{iso}$ (blue solid line), the resulting
    distribution for signal photons scaled so that the maximum value is the same 
    as that for identified photons (green dot-dashed line), and that for photons in
    simulation which are isolated at the generator level normalised to have the 
    same integral as the signal photon distribution (red dashed line). Each
    panel shows a different pseudorapidity region, while the top and bottom
    panels show the low-$E_\mathrm{T}^{\gamma}$ and
    high-$E_\mathrm{T}^{\gamma}$ range respectively. The 
    vertical error bars represent statistical uncertainties only.}
\label{fig:iso}
\end{center}
\end{figure}

Figure~\ref{fig:iso} shows example $E_\mathrm{T}^\mathrm{iso}$ 
distributions for identified and isolated photons, the corresponding distributions for 
background photons with the normalisation determined by the double-sideband 
method, and the resulting signal-photon distributions after background 
subtraction, compared with those for generator-level photons in MC simulation.
The figure shows the shape of the background distribution within the 
signal region, and the correspondence between the background-subtracted 
data and the signal-only \PYTHIA8 distributions gives
confidence that the simulations accurately represent the data.

\begin{figure}[t]
\begin{center}
\includegraphics[width=\textwidth]{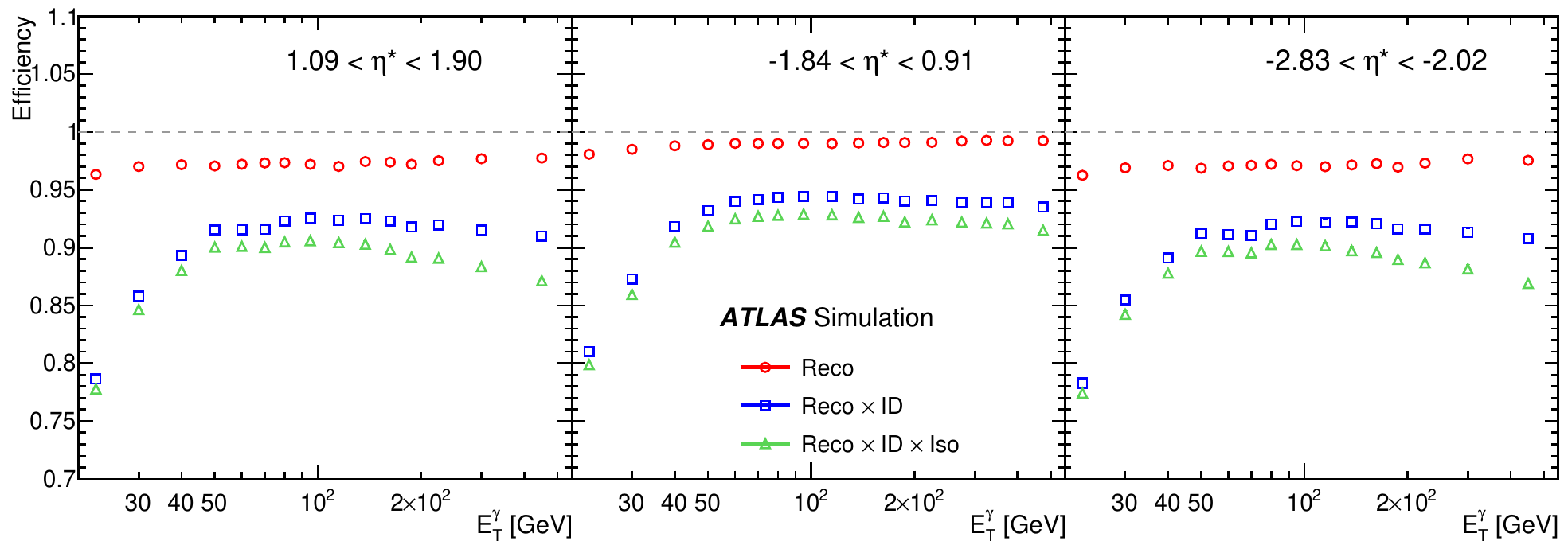}
\caption{Efficiency for simulated photons passing the
    generator-level isolation requirement, shown as a function of
    photon transverse energy $E_\mathrm{T}^{\gamma}$ 
    with a different pseudorapidity region
    in each panel. The reconstruction (red circles), reconstruction plus
    identification (blue squares) and total selection (green triangles) efficiencies are
    shown separately.
\label{fig:eff}}
\end{center}
\end{figure}

The photon selection efficiency is determined from MC simulations. 
Generated prompt photons are required to be isolated at the generator 
level, after an estimate of the underlying event has been subtracted from the isolation energy, 
as described above. Reconstruction efficiency is determined
by requiring a photon to have been reconstructed within $\Delta R=0.2$ of the
generated photon. Reconstructed photons matching to a generated photon are
further required to satisfy tight identification and isolation
criteria defined in Section~\ref{sec:intro}. The combined efficiency of signal
photons to pass all reconstruction level selections, $\epsilon^\mathrm{sel}$, 
is typically $90$\% at all $E_\mathrm{T}^{\gamma}$ and $\eta^*$, except at
$E_\mathrm{T}^{\gamma} \approx 20$~\GeV{} where it decreases to about
$80$\%.  Figure~\ref{fig:eff} summarises the different components of the
total selection efficiency. The reconstruction efficiency is 96--99\% 
everywhere, with the lowest values at the lowest $E_\mathrm{T}^{\gamma}$.
The isolation efficiency is lowest at high $E_\mathrm{T}^{\gamma}$, 
most likely because the associated products of fragmentation photons are,
on average, more energetic and collimated when the energy of the photon is 
higher. The largest inefficiency is due to the identification requirements.
This identification efficiency is lowest at 20~\GeV{} and increases with 
$E_\mathrm{T}^{\gamma}$ as higher-energy photons create larger and more 
identifiable showers in the calorimeter. It peaks around 100~\GeV{}, 
and decreases with increasing $E_\mathrm{T}^{\gamma}$ due to the 
difficulty of separating conversion electrons at high energy.

In MC events, the $E_\mathrm{T}$ response for prompt, isolated
photons, defined as the ratio of the reconstructed to generator
$E_\mathrm{T}$, is found to be within 1\% of unity, with a resolution
that decreases from $3$\% to $2$\% over the $E_\mathrm{T}^{\gamma}$
range of the measurement. The bin migration correction factors $C_{\mathrm{MC}}$ 
are determined using the event simulations described in Section~\ref{sec:MC}. 
They are defined as the bin-by-bin ratio 
$C_{\mathrm{MC}}=N_{\mathrm{MC}}^{\mathrm{part}}/N_{\mathrm{MC}}^{\mathrm{reco}}$ of the 
reconstructed, identified, and isolated photon $E_\mathrm{T}^{\gamma}$ spectrum, 
where $N_{\mathrm{MC}}^{\mathrm{part}}$ is the number in a given $E_\mathrm{T}^{\gamma}$ 
bin at the particle level and $N_{\mathrm{MC}}^{\mathrm{reco}}$ is the number in
the corresponding bin at the reconstruction level.

\begin{figure}[h]
\begin{center}
\includegraphics[width=0.6\textwidth]{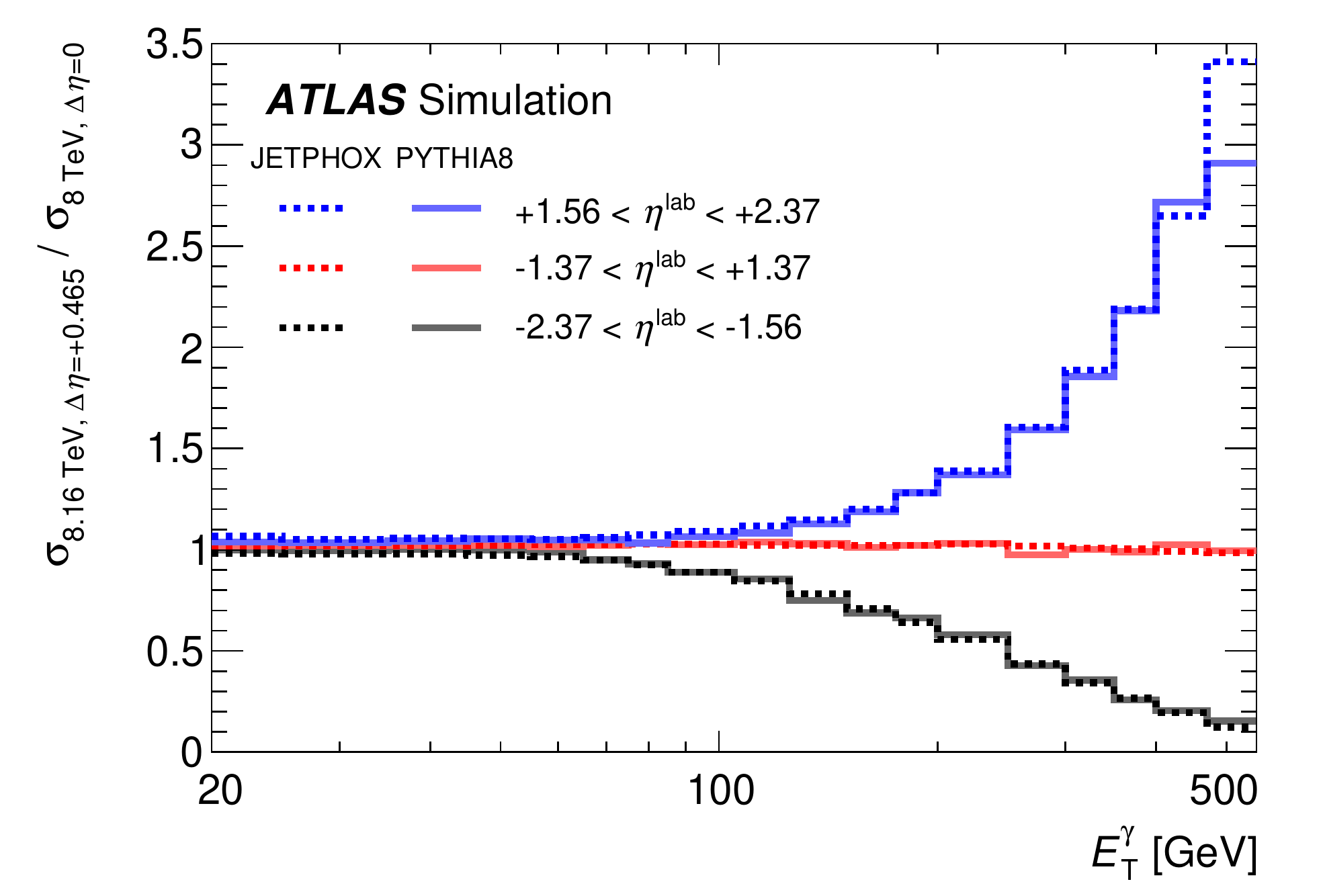}
\caption{Summary of extrapolation factors applied to the measured $pp$
  $\sqrt{s} = 8$~\TeV{} data to construct an approximate $\sqrt{s} =
  8.16$~\TeV{} spectrum matching the shift of the centre-of-mass in
  $p$+Pb data plotted as a function of generator-level photon 
  transverse energy. Here $\Delta \eta$ represents the boost of the centre-of-mass 
  frame of 0.465. The factors determined using \textsc{Jetphox} (dashed lines)
  and \PYTHIA8 (solid lines) are shown for the three
  $\eta^\mathrm{lab}$ ranges used in the measurement (different colours). 
  The relative difference between these two extrapolation methods is
  taken as a systematic uncertainty.
\label{fig:aux:0}}
\end{center}
\end{figure}

The nuclear modification factor $R_{p\mathrm{Pb}}$ can be expressed as a ratio 
of cross-sections in the following way:
\begin{equation}
R_{p\mathrm{Pb}} = ( \mathrm{d}\sigma^{p+\mathrm{Pb}\rightarrow\gamma+X}/ \mathrm{d}\ET^\gamma ) / ( A \cdot \mathrm{d}\sigma^{\pp\rightarrow\gamma+X}/ \mathrm{d}\ET^\gamma )\;, \label{eq:RpPbMB}
\end{equation}
where the geometric factor $A$ is simply the number of nucleons in the 
Pb nucleus, 208. The reference $pp$ spectrum is constructed using measurements of 
$\sqrt{s} = 8$~\TeV{} $pp$ data by ATLAS~\cite{STDM-2014-09} that
use the same particle-level isolation requirement. The 8~\TeV{} measurements in 
the regions $|\eta^\mathrm{lab}| < 1.37$ and $1.56 < |\eta^\mathrm{lab}| < 2.37$ 
are used as the reference spectra for the central and the forward and backward 
rapidity data, after applying a multiplicative correction for the effects of the boost 
in the 8.16~\TeV{} $p$+Pb system. For each kinematic region, extrapolation 
factors are determined as the ratio of photon cross sections from \textsc{Jetphox} 
calculations for $pp$ collisions. The numerator has $\sqrt{s}=8.16$~\TeV{} with a 
boost of the centre-of-mass corresponding to the $p$+Pb system, and the denominator 
has energy $\sqrt{s}=8$~\TeV{} with its rest frame corresponding with that of the 
laboratory reference frame. That is, the cross-sections in the numerator and 
denominator use the same $\eta^\mathrm{lab}$ regions, although in the former case 
this corresponds to a different center-of-mass pseudorapidity. These factors are 
shown in Figure~\ref{fig:aux:0} and are applied as multiplicative factors to the
measured $8$~\TeV{} data. They are dominated by the effect from the boost of the
$p$+Pb system, as the effect due to the difference in collision energy alone is less than $1$\% for
all $E_\mathrm{T}^{\gamma}$. For $-1.84 < \eta^* < 0.91$, or $E_\mathrm{T}^{\gamma} <
100$~\GeV{} at large rapidities, the factors are typically within a few
percent of unity. However, at large $E_\mathrm{T}^{\gamma}$, where the
rapidity distribution becomes steeper, the extrapolation factors
become more sensitive to the rapidity shift from the centre-of-mass 
boost between the frames, and at large pseudorapidity they reach
a factor of $2$--$3$. An alternative set of factors, derived from
the generator-level predictions of \PYTHIA8, are also shown in
Figure~\ref{fig:aux:0}; these are used to assess the sensitivity of the
extrapolation factors to the rapidity and $E_\mathrm{T}^{\gamma}$ dependence of 
the model cross sections.

\section{Systematic uncertainties}
\label{sec:systematic}

The sources of systematic uncertainties affecting the measurement are
described in this section, which is broken into two parts discussing
the uncertainty in 1) the cross-section and 2) the nuclear modification factor
$R_{p\mathrm{Pb}}$, including its ratio between forward and backward
pseudorapidity regions.

\subsection{Cross-section uncertainty}

\begin{figure}[t]
\begin{center}
\includegraphics[width=\textwidth]{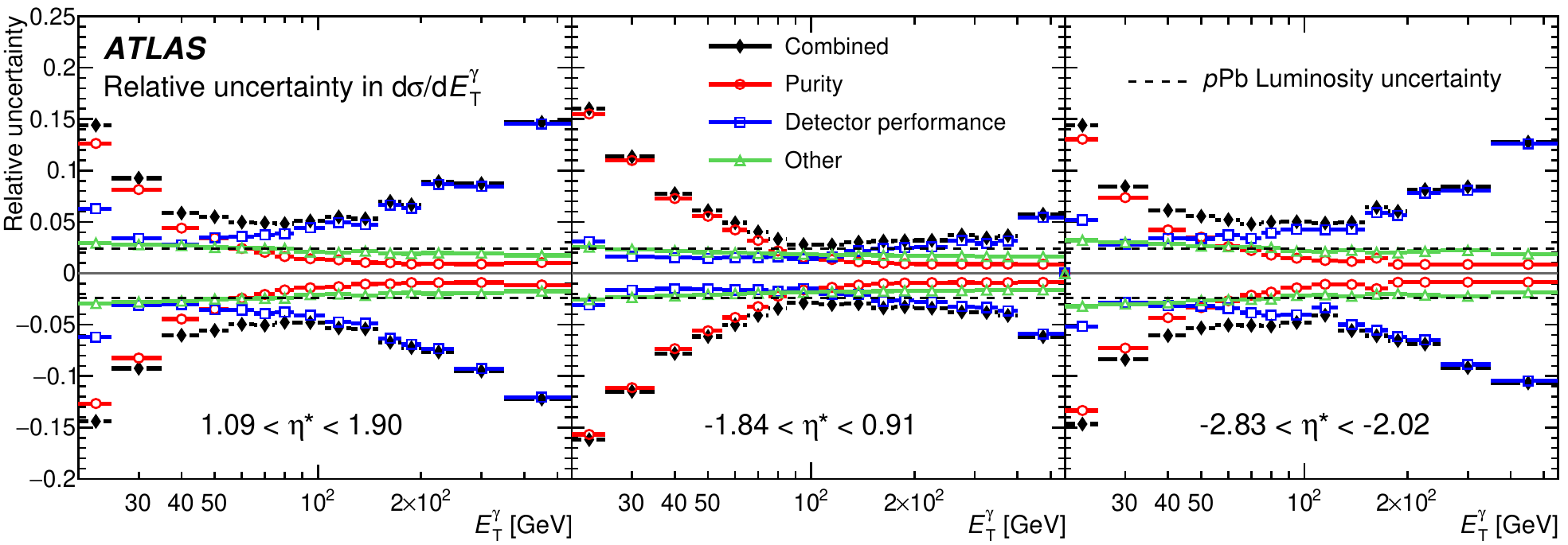}
\caption{Summary of the relative sizes of major sources of
    systematic uncertainty in the cross-section measurement, as well as the 
    combined uncertainty (excluding luminosity), shown
    as a function of photon transverse energy $E_\mathrm{T}^{\gamma}$.
\label{fig:syst}}
\end{center}
\end{figure}

The major uncertainties in the cross-section can be divided into two
main categories: those affecting the purity determination, which are 
dominant at low $E_\mathrm{T}^{\gamma}$ where the sample purity is low, and 
those affecting the detector performance corrections, which are dominant at 
high $E_\mathrm{T}^{\gamma}$. 
All other sources tend to be weakly dependent on $E_\mathrm{T}^{\gamma}$. 
A summary is shown in Figure~\ref{fig:syst}. In each category, the uncertainty is the
sum in quadrature of the individual components; the combined uncertainty is the sum 
in quadrature of all contributions, excluding those associated with the luminosity. 
The total uncertainties range from 15\% at low
and high $E_\mathrm{T}^{\gamma}$, where they are dominated by the
purity and detector performance uncertainties respectively, to a
minimum of approximately $6$\% at $E_\mathrm{T}^{\gamma} \approx 100$~\GeV,
where both of these uncertainties are modest. 

To assess the uncertainty in the purity determination, each boundary defining 
the sidebands used in the calculation is varied independently in order to
understand the sensitivity of the measurement to the double sideband binning and correlation assumptions. 
The dominant uncertainty arises from uncertainty in the level of sideband 
correlation. This is estimated directly from data by dividing the non-isolated
region in two subregions and calculating the ratio of identified to 
background-enhanced yields in each subregion. These ratios differ at the level of
10\% which agrees with estimates from previous studies~\cite{STDM-2016-08}.
This $\pm10\%$ variation in the sideband correlation yields a 13\% 
uncertainty in the cross-section in the lowest $E_\mathrm{T}^{\gamma}$
range, decreasing to less than $1$\% for $E_\mathrm{T}^{\gamma} > 100$~\GeV{}.
The inverted photon identification requirement for the background candidates is varied to be less or more 
restrictive about which shower shapes the background candidates are required 
to fail. This variation yields an uncertainty that is less than $1$\% for all $E_\mathrm{T}^{\gamma}$ 
in the forward and backward rapidity bins, but is significant at mid-rapidity
($-1.84 < \eta^* < 0.91$) where it is 9\% in the lowest $E_\mathrm{T}^{\gamma}$ bin and decreases 
to be less than $1$\% for $E_\mathrm{T}^{\gamma} > 100$~\GeV{}. Variations in the 
isolation energy threshold of $\pm 1$~\GeV{} have been shown to cover any difference 
between simulations and data~\cite{STDM-2016-08}. These variations result in a 1--2\% effect on the
cross-section in the lowest $E_\mathrm{T}^{\gamma}$ range and less than $1$\% at 
higher $E_\mathrm{T}^{\gamma}$. The uncertainty associated with the inverted
shower-shape was smoothed and symmetrised, however, the other uncertainties
are derived asymmetrically from the positive and negative variations separately.

Uncertainties associated with detector performance corrections are dominant 
at high $E_\mathrm{T}^{\gamma}$. A detailed description of the several 
components of the photon energy scale and resolution uncertainties are given 
in Ref.~\cite{STDM-2016-08}. The impact of these on the measurement is
determined by varying the reconstructed photon $E_\mathrm{T}^{\gamma}$ in 
simulation within the energy scale uncertainties and deriving alternative correction factors for positive and negative 
variations separately. Of these, the impact of the energy scale variation is dominant, giving a 
10--15\% contribution at 500~\GeV{} in the forward and backward regions, 
decreasing to less than $2$\% at the lowest $E_\mathrm{T}^{\gamma}$. In the mid-rapidity 
region, the energy scale variation gives a 5\% uncertainty at high $E_\mathrm{T}^{\gamma}$, decreasing to 
less than $1$\% at low $E_\mathrm{T}^{\gamma}$. Additionally, there are uncertainties associated 
with corrections for small differences in reconstruction, identification and isolation 
efficiencies observed between data and simulation~\cite{PERF-2017-02}. These uncertainties 
are about 5\% in the forward regions and low $E_\mathrm{T}^{\gamma}$ and less 
than $2$\% elsewhere.

Systematic uncertainties related to modelling in simulation,
luminosity, electron contamination, and other sources tend to be lower 
than those previously discussed. However, their combined effect is 
dominant in the mid-rapidity region and between 90~\GeV{} and 
250~\GeV{}. To test the sensitivity of the measurement to the difference 
of isolation energy between particle-level and detector level 
in the simulation, the generator-level isolation definition
is changed to better correspond to the reconstruction-level
definition. The relative change in the cross-section after this
deviation from the nominal is about 2\% at low $E_\mathrm{T}^{\gamma}$,
decreasing to about 1\% at high $E_\mathrm{T}^{\gamma}$, for
each pseudorapidity region, and is taken as a symmetric uncertainty. An 
uncertainty is assigned to cover the possible contribution of
misreconstructed electrons, primarily from the decays of $W^\pm$
and $Z$ bosons, to the selected photon
yield. Based on simulation studies, and the results of previous
measurements~\cite{STDM-2014-09,STDM-2016-08}, this is assigned to be 
1.3\% for $E_\mathrm{T}^{\gamma} < 105$~\GeV{} in forward 
pseudorapidity regions, and 0.5\% everywhere else. To test the beam 
orientation dependence, the cross-section is measured 
using the data from each beam configuration separately. The two 
measurements agree at the level of 1\%, well above the statistical 
uncertainty for most $E_\mathrm{T}^{\gamma}$ bins. This difference 
is taken as a global, symmetric uncertainty in the combined results.
To test the sensitivity to the relative fractions of direct and fragmentation 
photons in the MC samples, the simulation is weighted such that the fraction 
of direct photons is unity, that is, all photons in the sample are direct. This
reflects a conservative difference compared with the default estimate of this
fraction of about $50-80$\% from the MC samples. This variation
gives a relative change in the cross-section of approximately 1\% for all 
kinematic regions, which is taken as a systematic uncertainty. The uncertainty 
in the integrated luminosity of the combined data sample is $2.4$\% It is 
derived, following a methodology similar to that detailed in 
Ref.~\cite{DAPR-2013-01}, and using the LUCID-2 detector for the baseline 
luminosity measurements \cite{LUCID2}, from calibration of the luminosity 
scale using x-y beam-separation scans.

\subsection{$R_{p\mathrm{Pb}}$ uncertainty}
\label{sec:sys:RpPb}

The nuclear modification factor $R_{p\mathrm{Pb}}$ is affected by systematic uncertainties 
associated with the $p$+Pb and $pp$ measurements. The uncertainties in the differential 
cross-section of the $pp$ reference data are obtained directly from Ref.~\cite{STDM-2014-09}.
Due to differences in photon reconstruction, energy 
calibration, isolation and identification procedures between the $pp$ and $p$+Pb
datasets, the uncertainties are treated as uncorrelated and added in 
quadrature.

The uncertainty in the extrapolation of the $pp$ $E_\mathrm{T}^{\gamma}$ 
spectrum at $8$~\TeV{} is determined by using an alternative method to derive the 
multiplicative extrapolation factors. Instead of \textsc{Jetphox}, photon
cross-sections for the $8$~\TeV{} and rapidity-boosted $\sqrt{s} = 8.16$~\TeV{} 
kinematics are determined from \PYTHIA8. The extrapolation factors from both 
\textsc{Jetphox} and \PYTHIA8 are shown in Figure \ref{fig:aux:0}.
Additionally, \textsc{Jetphox} is run with an alternative PDF set to quantify the 
impact of a given PDF choice. The differences between the extrapolation factors from 
these two variations, which are at most a few percent in the kinematic region of the 
measurement and subdominant with respect to the other uncertainties in the cross-sections, 
are added in quadrature and used as an estimate of the uncertainty in the 
extrapolation procedure.

For the measurement of the ratio of $R_{p\mathrm{Pb}}$ values (Eq. (\ref{eq:RpPbMB})) 
between the forward and backward pseudorapidity regions, each 
systematic variation affecting the purity and detector performance corrections 
is applied to the numerator and denominator in a coherent way, allowing 
them to partially cancel out in the ratio. All uncertainties in the other categories, 
except those from electron contamination and the beam direction difference, 
are treated as correlated. For this reason, they cancel out; notably the $p$+Pb luminosity 
and $pp$ cross-section uncertainties cancel out completely. The extrapolation 
uncertainties are treated as independent and are added in quadrature to
the other uncertainties in $R_{p{\mathrm Pb}}$.

\begin{figure}[t]
\begin{center}
\includegraphics[width=0.60\textwidth]{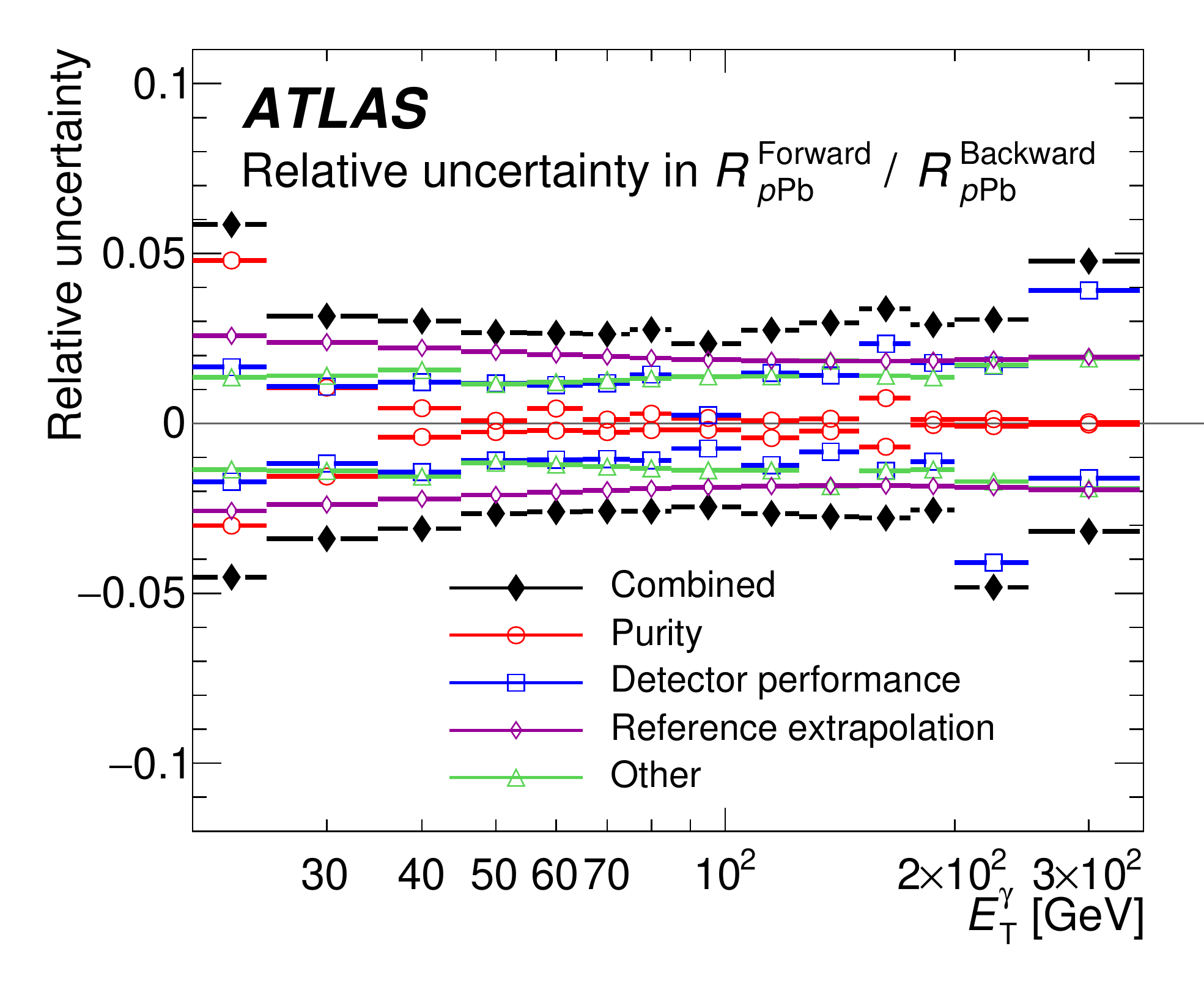}
\caption{ Summary of the relative size of major sources of systematic 
    uncertainty in the forward-to-backward ratio of the nuclear modification
    factor $R_{p\mathrm{Pb}}$, as well as the combined uncertainty, shown as 
    a function of photon transverse energy $E_\mathrm{T}^{\gamma}$. The Reference 
    extrapolation refers to the uncertainty related to the extrapolation of the 
    previously measured $8$~\TeV{} $pp$ spectrum to $8.16$~\TeV{} and boosted
    kinematics.
\label{fig:sysfb}}
\end{center}
\end{figure}

The resulting uncertainty ranges from about 5\% at the lowest 
$E_\mathrm{T}^{\gamma}$, where it is dominated by the uncertainty 
in the purity, to about 3\% at mid-$E_\mathrm{T}^{\gamma}$, and again
about 5\% at high $E_\mathrm{T}^{\gamma}$, where it is dominated by
uncertainty in the energy scale. A summary of the uncertainties
in the forward-to-backward ratio is shown in Figure~\ref{fig:sysfb}.

\section{Results}
\label{sec:result}

\begin{figure}[t]
\begin{center}
\includegraphics[width=\textwidth]{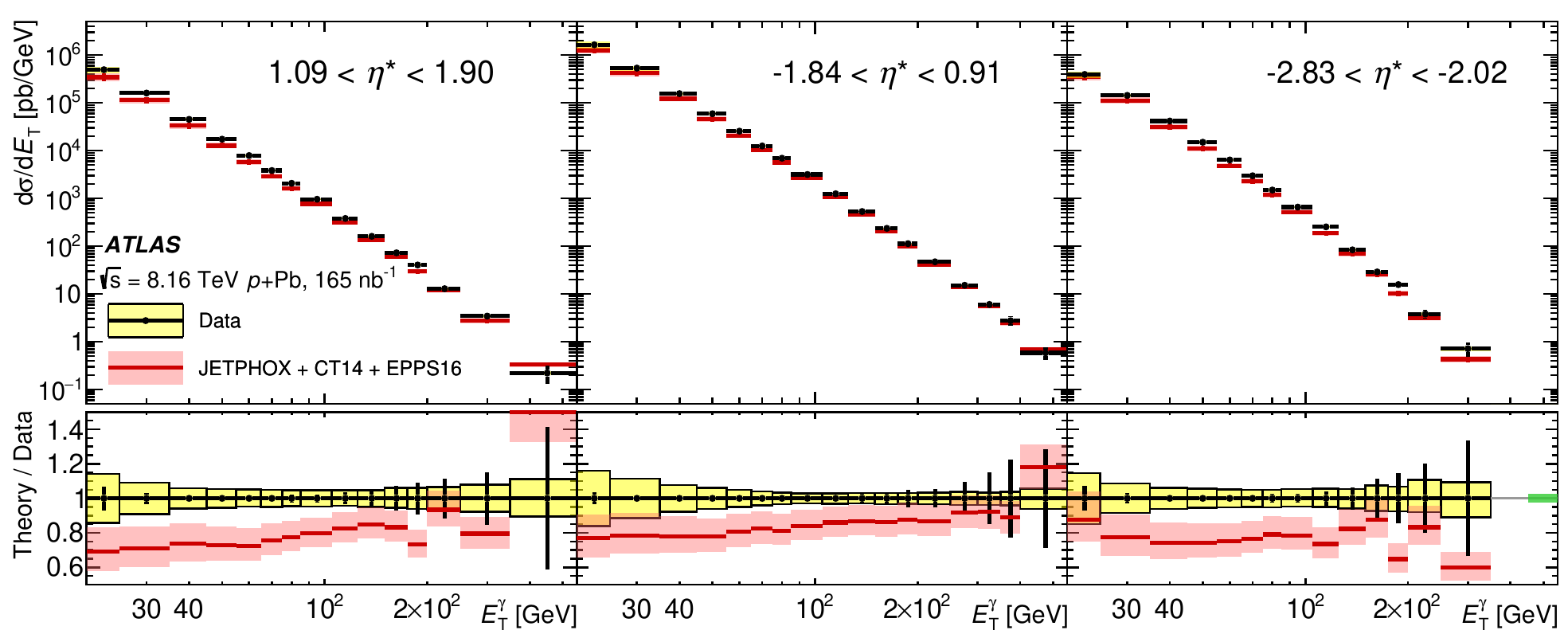}
\caption{Prompt, isolated photon cross-sections as a function of
  transverse energy $E_\mathrm{T}^{\gamma}$, shown for different centre-of-mass 
  pseudorapidity, $\eta^*$, regions in each panel. The data are compared with
  \textsc{Jetphox} with the EPPS16 nuclear PDF set~\cite{Eskola:2016oht}, 
  with the ratio of theory to data shown in the lower panels.
  Yellow bands correspond to total systematic uncertainties in
  the data (not including the luminosity uncertainty), vertical bars correspond 
  to the statistical uncertainties in the data, and the red bands correspond to the uncertainties in
  the theoretical calculation. The green box (at the far right) represents the 2.4\%
  luminosity uncertainty.
\label{fig:0}}
\end{center}
\end{figure}

Photon production cross-sections are shown in Figure~\ref{fig:0} for 
photons with $E_\mathrm{T}^{\gamma} > 20$~\GeV{} in three
pseudorapidity regions. The measured 
$\mathrm{d}\sigma/\mathrm{d}E_\mathrm{T}^{\gamma}$ values decrease 
by five orders of magnitude over the complete $E_\mathrm{T}^{\gamma}$ 
range, which extends out to $E_\mathrm{T}^{\gamma} \approx 
500$~\GeV{} for photons at mid-rapidity. In \PYTHIA8, photons in this
range typically arise from parton configurations in which the parton
in the nucleus has Bjorken scale variable, $x_\mathrm{A}$, in the range 
$3\times 10^{-3}\lesssim x_\mathrm{A}\lesssim 4\times 10^{-1}$. In the
nuclear modified PDF (nPDF) picture, this range probes the so-called 
shadowing (suppression for $x_\mathrm{A} \lesssim 0.1$), anti-shadowing 
(enhancement for $ 0.1 \lesssim x_\mathrm{A} \lesssim 0.3$), and EMC 
(suppression for $ 0.3 \lesssim x_\mathrm{A} \lesssim 0.7$) 
regions~\cite{Eskola:2016oht}.

The data are compared with an NLO pQCD calculation similar to that 
used in Ref.~\cite{Helenius:2014qla}, where the data is similarly underestimated at low ET, but using the updated
CT14~\cite{Dulat:2015mca} PDF set for the free-nucleon parton
densities. \textsc{Jetphox}~\cite{Aurenche:2006vj} is used to perform
a full NLO pQCD calculation of the direct and fragmentation
contributions to the cross-section. The BFG set
II~\cite{Bourhis:1997yu} of parton-to-photon fragmentation functions
are used, the number of massless quark flavours is set to five, and
the renormalisation, factorisation and fragmentation scales are chosen
to be $\ET^\gamma$. In addition to the calculation with the free-nucleon
PDFs, separate calculations are performed with the
EPPS16~\cite{Eskola:2016oht} and nCTEQ15~\cite{Kovarik:2015cma}
nPDF sets. The EPPS16 calculation uses the same
free-proton PDF set, CT14, as the free-nucleon baseline to which the
modifications are applied. The prediction is systematically lower than 
the data by up to 20\% at low $E_\mathrm{T}^{\gamma}$ but 
is closer to the data at higher $E_\mathrm{T}^{\gamma}$, consistent with 
the results of such comparisons in $pp$ collisions at LHC 
energies~\cite{STDM-2014-09,STDM-2016-08}. A recent calculation of isolated 
photon production at NNLO found that the predicted cross-sections were 
systematically larger at low $E_\mathrm{T}^{\gamma}$ than the NLO 
prediction~\cite{Chen:2019zmr}, and thus may provide a better description 
of the data in this and previous measurements.

\begin{figure}[h]
  \centering
  \includegraphics[width=.98\linewidth]{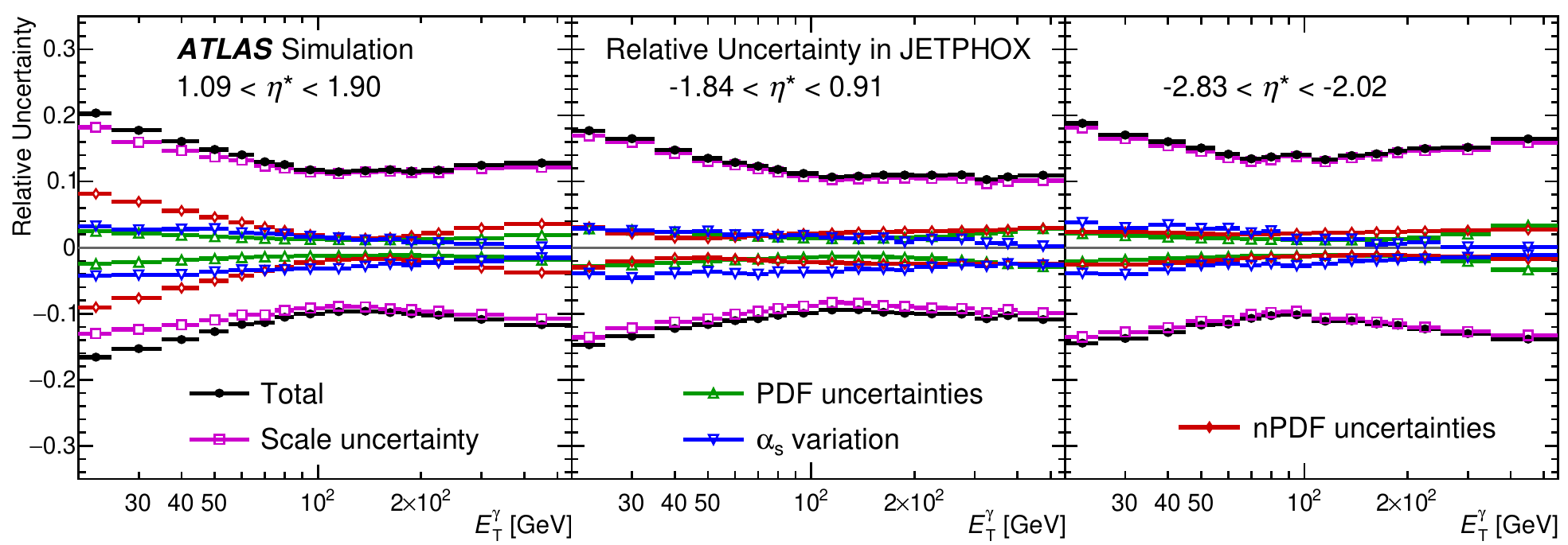}
  \caption{A breakdown of all systematic uncertainties in the cross-section
  prediction from \textsc{Jetphox} with the EPPS16 nPDF set.}
  \label{fig:theorySys} 
\end{figure}

Uncertainties associated with these calculations are assessed in a number 
of ways. Factorisation, renormalisation, and fragmentation scales are varied, 
up and down, by a factor of two as in 
Ref~\cite{STDM-2014-09}. The uncertainty is taken as the envelope formed by 
the minimum and maximum of each variation in every kinematic region and is 
dominant in most regions. PDF uncertainties are calculated via the standard 
CT14 error sets and correspond to a 68\% confidence interval. Again following 
Ref~\cite{STDM-2014-09} the sensitivity to the choice of $\alphas$ is 
evaluated by varying $\alphas$ by $\pm0.002$ around the central value 
of 0.118 in the calculation and PDF. Uncertainties from nPDFs 
are calculated from the error sets which correspond to 90\% confidence 
intervals, as described in Ref.~\cite{Eskola:2016oht}. These are converted into 
uncertainty bands which correspond to a 68\% confidence interval. A summary of each variation 
is shown in Figure~\ref{fig:theorySys}.

\begin{figure}[p]
\begin{center}
\includegraphics[width=\textwidth]{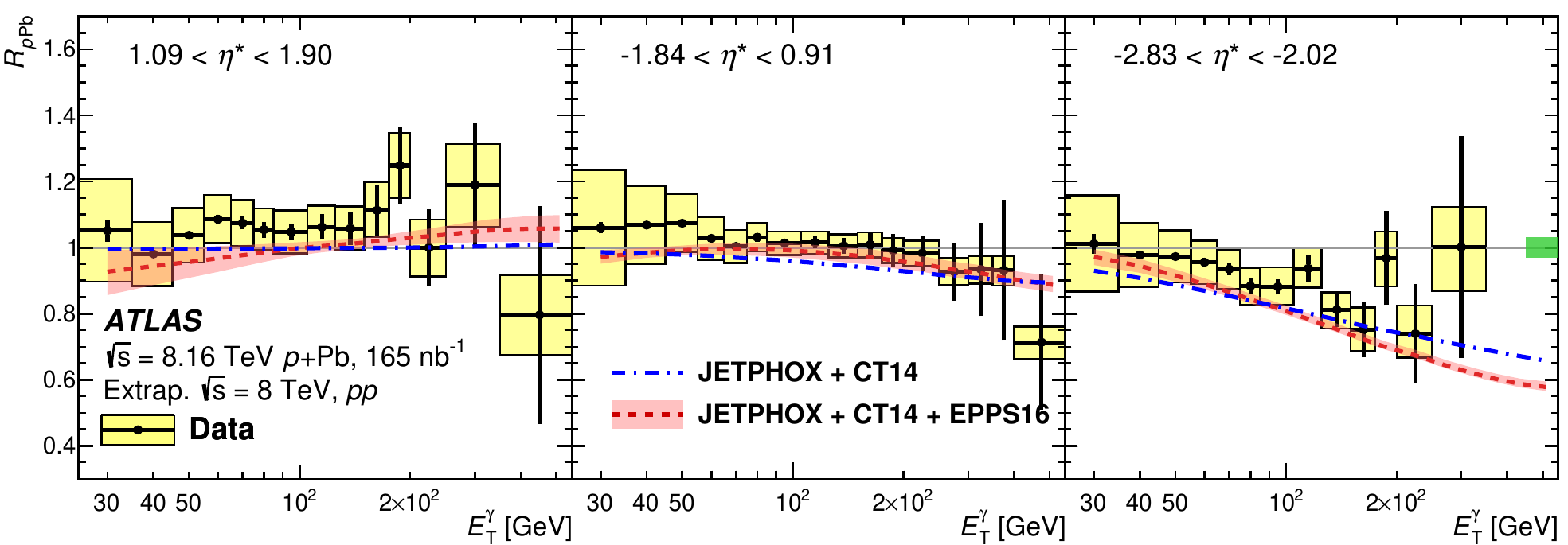}
\includegraphics[width=\textwidth]{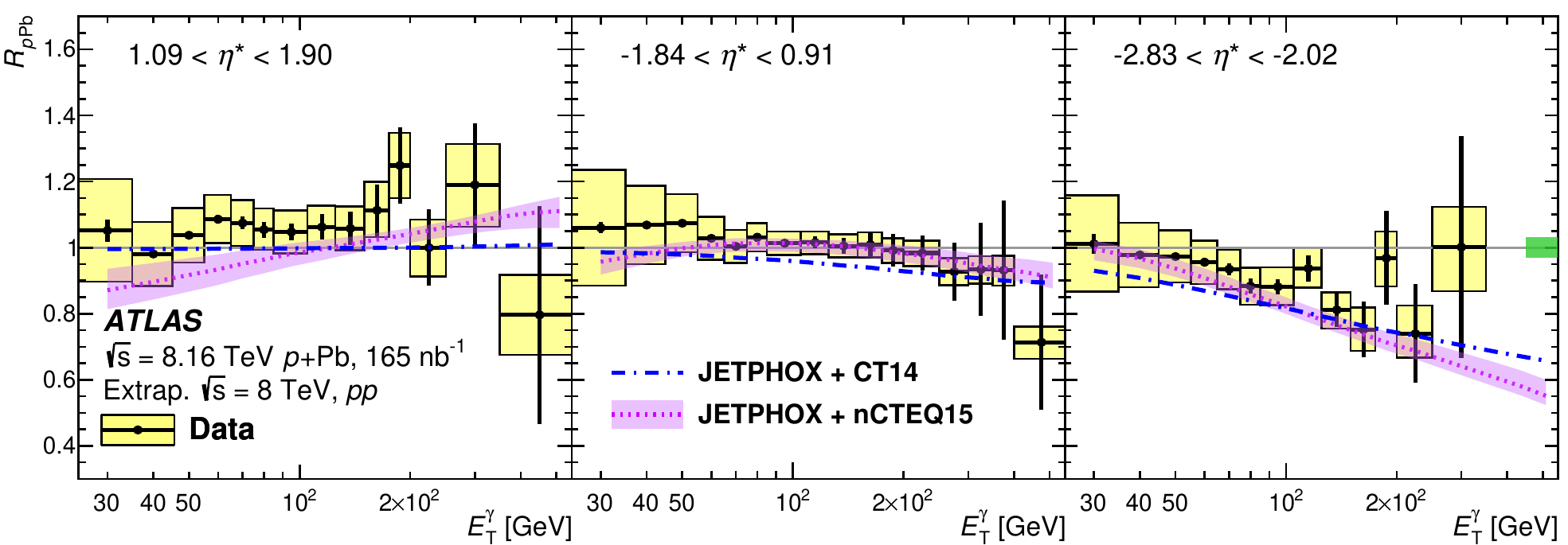}
\includegraphics[width=\textwidth]{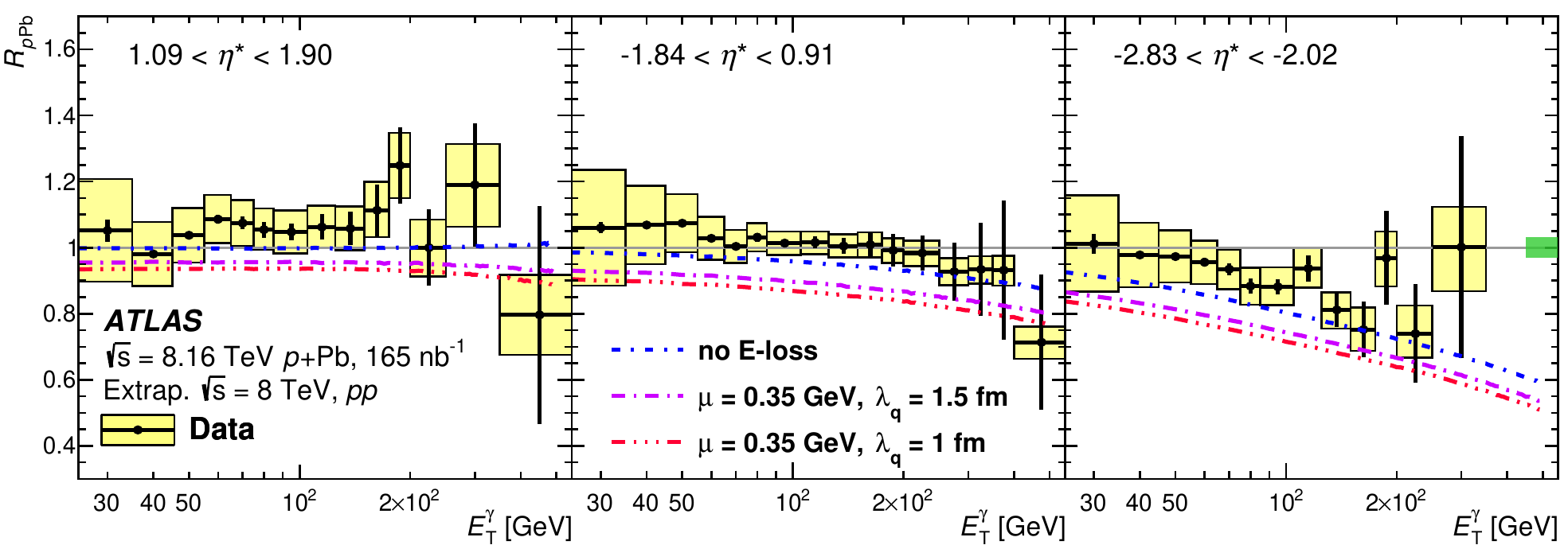}
\caption{Nuclear modification factor $R_{p\mathrm{Pb}}$ for isolated,
  prompt photons as a function of photon transverse energy
  $E_\mathrm{T}^{\gamma}$, shown for different centre-of-mass
  pseudorapidity, $\eta^*$, regions in each panel.  The
  $R_{p\mathrm{Pb}}$ is measured using a reference which is a
  simulation-derived extrapolation from $\sqrt{s} = 8$~\TeV{} $pp$
  data (see text).  The data are identical in each row, but show
  comparisons with the expectations based on \textsc{Jetphox} with the
  EPPS16 nuclear PDF set (top)~\cite{Eskola:2016oht}, with the nCTEQ15
  nuclear PDF set (middle)~\cite{Kovarik:2015cma}, and with an
  initial-state energy-loss calculation
  (bottom)~\cite{Vitev:2008vk,Kang:2015mta,Chien:2015vja}.  In all
  plots, the yellow bands and vertical bars correspond to total
  systematic and statistical uncertainties in the data
  respectively. In the top and middle panels, the red and purple bands
  correspond to the systematic uncertainties in the theoretical
  calculations. The green box (at the far right) represents the
  combined 2.4\% $p$+Pb and 1.9\% $pp$ luminosity uncertainties.
\label{fig:1}}
\end{center}
\end{figure}

Figure~\ref{fig:1} shows the nuclear modification factor
$R_{p\mathrm{Pb}}$ as a function of $E_\mathrm{T}^{\gamma}$ in different
$\eta^*$ regions. At forward rapidities ($1.09 < \eta^* < 1.90$), 
the $R_{p\mathrm{Pb}}$ value is consistent 
with unity, indicating that nuclear effects are small. In \PYTHIA8,
photons in this region typically arise from configurations with gluon 
partons from the Pb nucleus with $x_\mathrm{A} \approx 10^{-2}$. Nuclear modification 
pulls the pQCD calculation down slightly for 
$E_\mathrm{T}^{\gamma} < 100$~\GeV{}, above which the modification 
reverses, indicating a crossover between shadowing and 
anti-shadowing regions. At mid-rapidity, nuclear effects are similarly small
and consistent with unity at low $E_\mathrm{T}^{\gamma}$, but at higher 
$E_\mathrm{T}^{\gamma}$, there is a hint that $R_{p\mathrm{Pb}}$ is lower. This 
feature primarily reflects the different up- and down-quark composition
of the nucleus relative to the proton and is more important at larger parton $x$. 
In this case, the larger relative down-quark density decreases the photon yield. This effect is evident in the 
\textsc{Jetphox} theory curve in blue dash-dotted line, which includes the proton--neutron 
asymmetry and the free-nucleon PDF set CT14. This effect is most
pronounced at backward pseudorapidity where, in \PYTHIA8, the nuclear 
parton composition is typically a quark with $x_\mathrm{A}\approx 0.2$. Here, nPDF 
modification moves $R_{p\mathrm{Pb}}$ above the free-nucleon PDF 
calculation at low $E_\mathrm{T}^{\gamma}$ but below at high 
$E_\mathrm{T}^{\gamma}$, indicating the crossover from the anti-shadowing 
to the EMC region. 

The $R_{p\mathrm{Pb}}$ calculations including nPDFs consider only the 
nPDF uncertainty, since previous calculations have shown that the scale 
and PDF uncertainties cancel out almost completely in the kinematic region 
of the measurement~\cite{Helenius:2014qla}, and no non-perturbative 
corrections are applied. Within the present uncertainties, the data are 
consistent with both the free-proton PDFs and with the small effects expected 
from a nuclear modification of the parton densities. 

The $R_{p\mathrm{Pb}}$ measurements are also compared with an initial-state energy-loss 
prediction that is calculated within the framework described in
Refs.~\cite{Vitev:2008vk,Kang:2015mta,Chien:2015vja}. In this model,
the energetic partons undergo multiple scattering in the cold nuclear
medium, and thus lose energy due to this medium-induced gluon
bremsstrahlung, before the hard collision. The calculation is
performed with a parton--gluon momentum transfer $\mu = 0.35$~\GeV{}
and mean free path for quarks $\lambda_q = 1.5$~fm. Alternative
calculations with a shorter path length ($\lambda_q = 1$~fm), and a
control version with no initial-state energy loss, are also considered.
The data disfavour a large suppression of the cross-section from
initial-state energy-loss effects.

\begin{figure}[t]
\begin{center}
\includegraphics[width=0.49\textwidth]{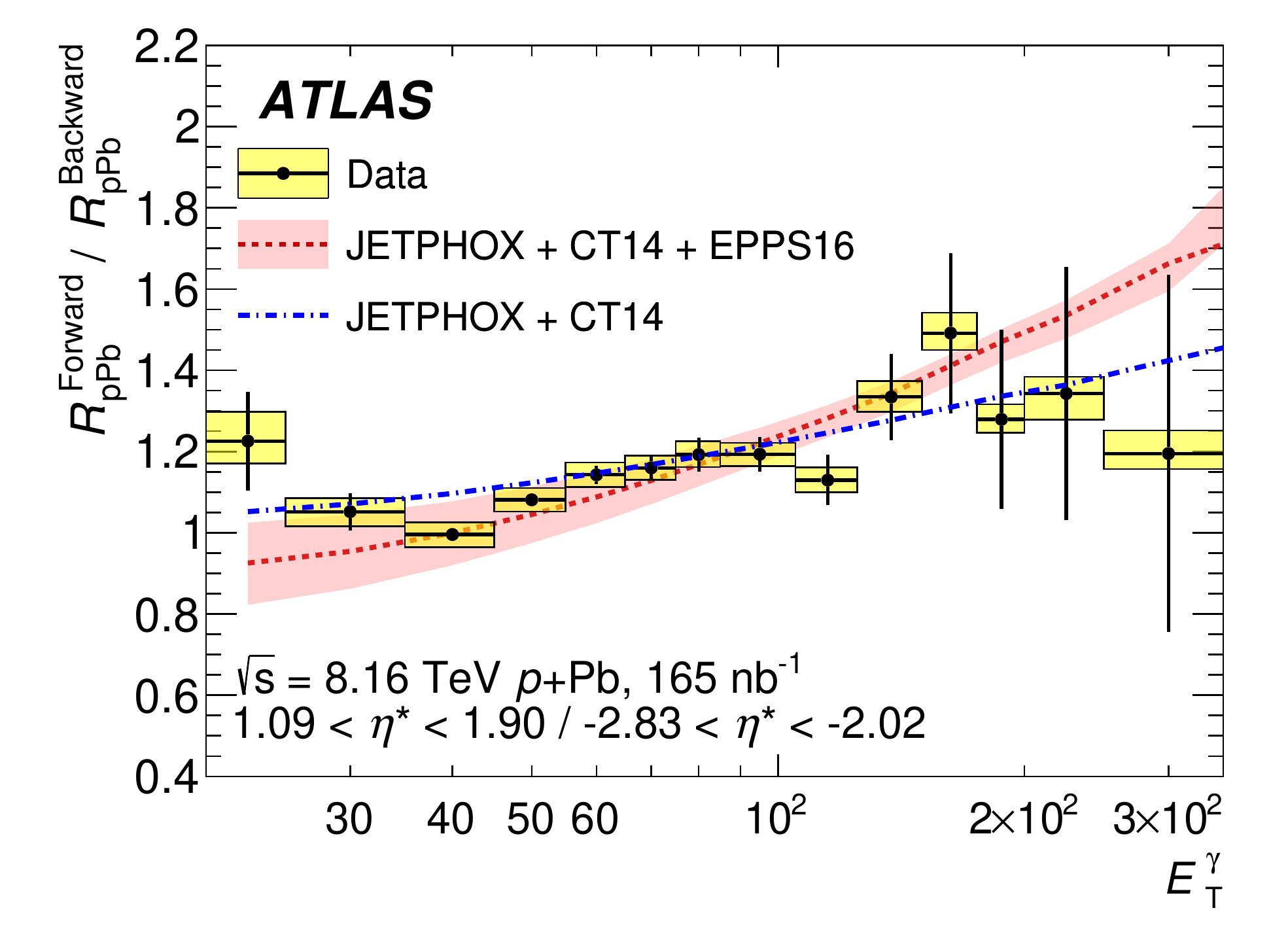}
\includegraphics[width=0.49\textwidth]{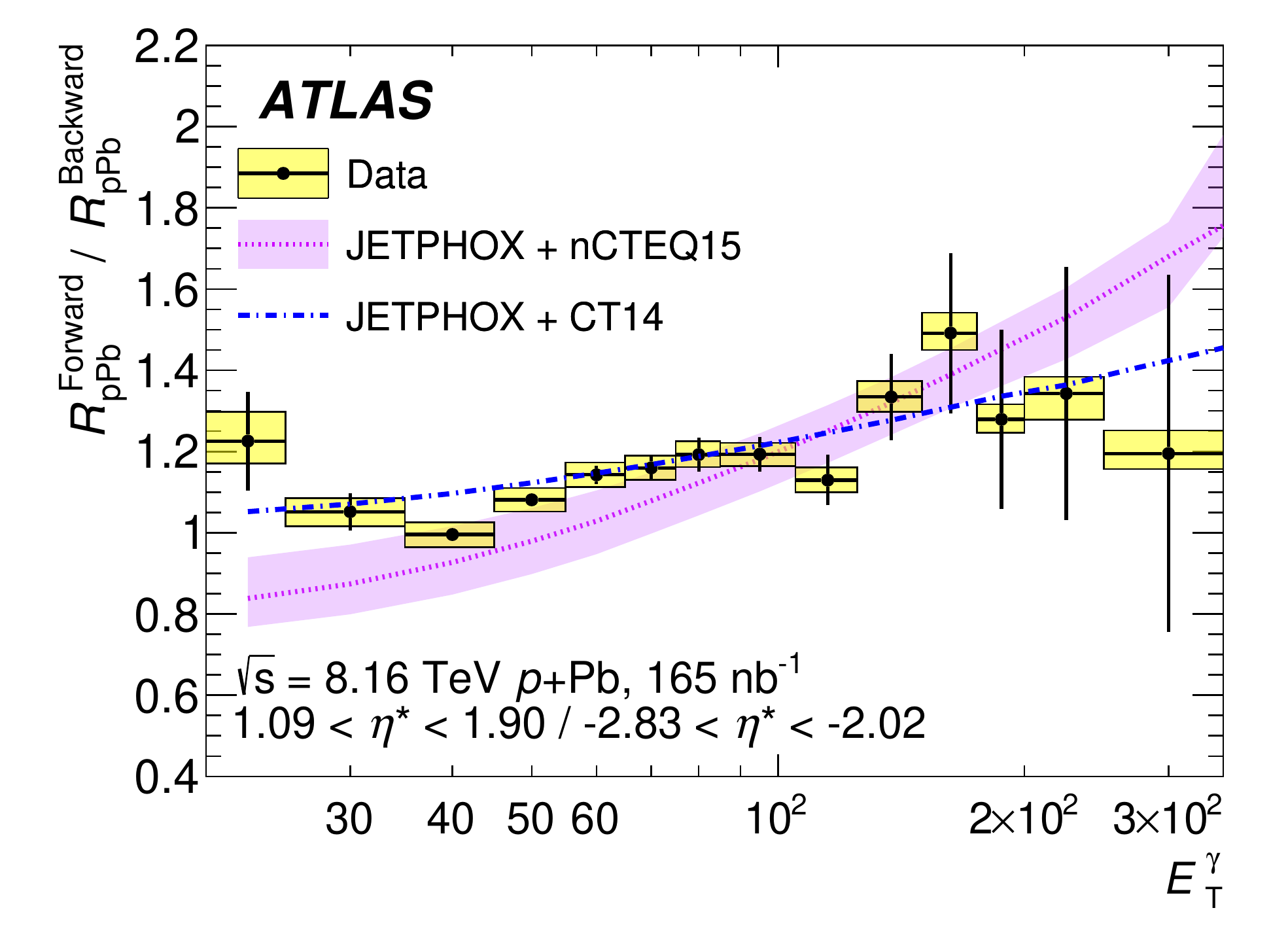}
\includegraphics[width=0.49\textwidth]{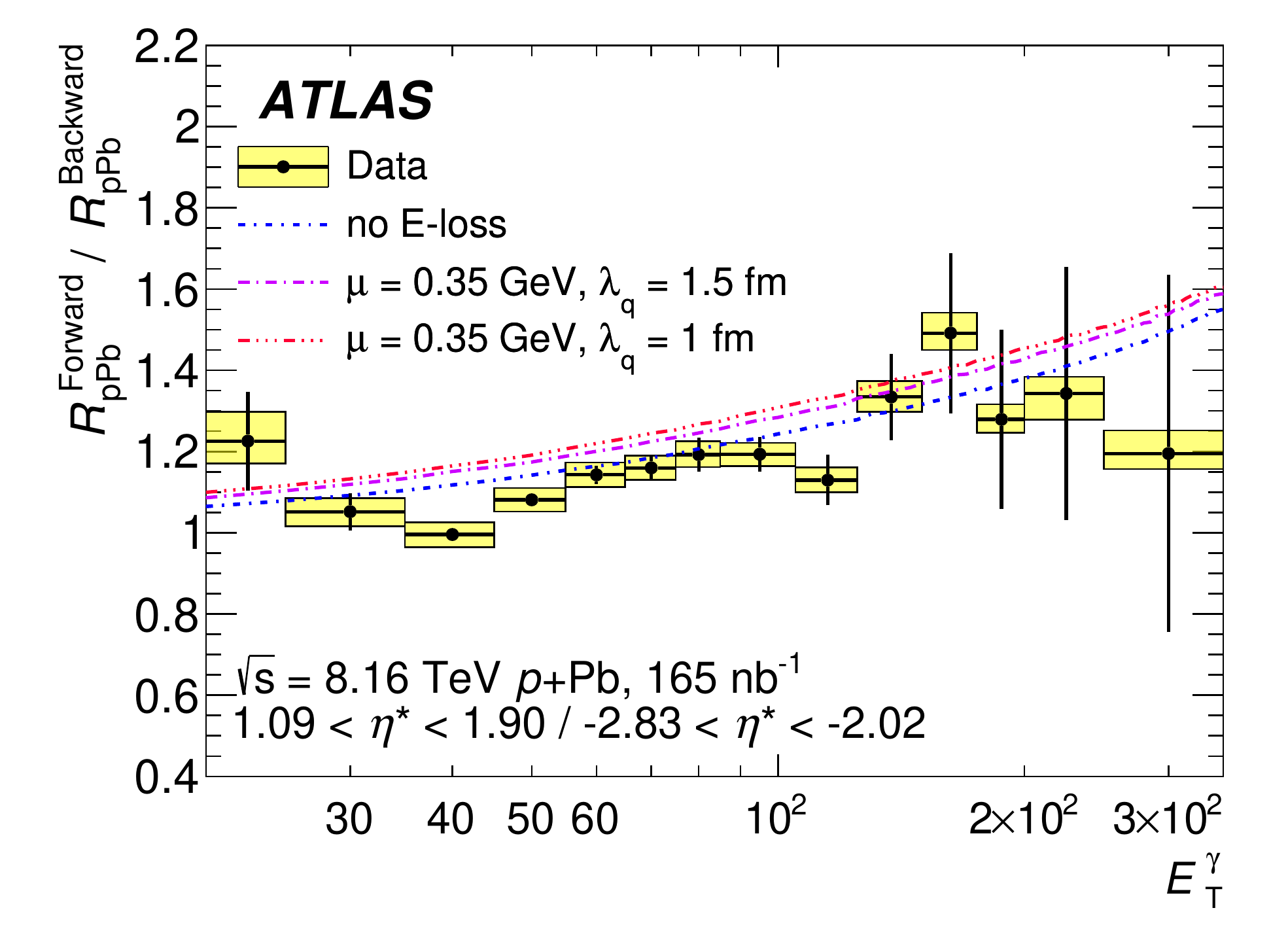}
\caption{Ratio of the nuclear modification factor
    $R_{p\mathrm{Pb}}$ between forward and backward pseudorapidity for
    isolated, prompt photons as a function of photon transverse energy
    $E_\mathrm{T}^{\gamma}$. The data are identical in each panel, but
    show comparisons with the expectations based on \textsc{Jetphox}
    with the EPPS16 nuclear PDF set (top, left)~\cite{Eskola:2016oht} or with the nCTEQ15 nuclear
    PDF set (top, right)~\cite{Kovarik:2015cma}, and with an initial-state energy-loss calculation
    (bottom)~\cite{Vitev:2008vk,Kang:2015mta,Chien:2015vja}. The strength
    of the initial-state energy-loss effect is parameterised by $\lambda_\textrm{q}$,
    which represents the mean free path of partons in the nuclear medium and 
    is smaller for a larger degree of energy loss. In all plots, the yellow 
    bands and vertical bars correspond to total systematic and statistical 
    uncertainties in the data respectively. In the left and right panels, 
    the red and purple bands correspond to the systematic uncertainties in the
    calculations.
\label{fig:2}}
\end{center}
\end{figure}

The ratio of the $R_{p\mathrm{Pb}}$ values between forward and backward
pseudorapidity, shown in Figure~\ref{fig:2}, is studied as a way to
reduce the effect of common systematic uncertainties and better isolate the 
magnitude of nuclear effects~\cite{Paukkunen:2010qg}. The remaining 
systematic uncertainty, discussed in Sec.~\ref{sec:sys:RpPb}, is dominated 
by the reference extrapolation and treated as uncorrelated between points. 
Below $E_\mathrm{T}^{\gamma}\approx 100$~\GeV{}, this corresponds 
roughly to the ratio of $R_{p\mathrm{Pb}}$ from photons from gluon 
nuclear parton configurations in the shadowing $x_\mathrm{A}$ region to that from quark  
partons in the anti-shadowing region. This can be seen in the top two 
panels of Figure~\ref{fig:2}, where the nuclear modification (red/purple 
bands) brings the \textsc{Jetphox} calculation below that of the free-nucleon 
PDF (blue curve), though the effect from EPPS16 is less significant. 
In contrast, the behaviour is reversed at higher $E_\mathrm{T}^{\gamma}$ 
where the numerator probes the shadowing/anti-shadowing crossover 
region and the denominator moves deeper into the EMC region~\cite{Eskola:2016oht}. 
The data are consistent with the pQCD calculation before incorporating 
nuclear effects, except possibly in the region $E_\mathrm{T}^{\gamma} < 55$~\GeV, 
which is sensitive to the effects from gluon shadowing. At low $E_\mathrm{T}^{\gamma}$, the data are 
systematically higher than the calculations which incorporate nPDF effects, 
but approximately within their theoretical uncertainty. Additionally, in the
lower plot of Figure~\ref{fig:2}, the forward-to-backward ratios are 
compared with predictions from a model incorporating initial-state energy 
loss. The data show a preference for no or only a limited amount of 
energy loss.

\section{Conclusion}
\label{sec:conclusion}

This letter presents a measurement of the inclusive prompt, isolated 
photon cross-section in $p$+Pb collisions at $\sqrt{s_\mathrm{NN}} =
8.16$~\TeV, using a dataset corresponding to an integrated luminosity
of $165$~nb$^{-1}$ recorded by the ATLAS experiment at the LHC.
The differential cross-section as a function of the photon transverse 
energy is reported in three pseudorapidity regions in the 
nucleon--nucleon collision frame, and covers photon transverse 
energies from 20~\GeV\ to 550~\GeV. The data are compared with a next-to-leading-order
calculation which incorporates nuclear PDF effects. A measurement of
the nuclear modification factor is reported in the region above 25~\GeV\ using 
a NLO pQCD-based extrapolation of previously published $pp$ data at 
$\sqrt{s} = 8$~\TeV. The data are compatible with the expectation that the 
PDFs are modestly modified in nuclei in this kinematic region and may help to 
place an upper limit on the possible amount of energy lost by partons in the initial 
stages of nuclear collisions.

\section*{Acknowledgements}


We thank CERN for the very successful operation of the LHC, as well as the
support staff from our institutions without whom ATLAS could not be
operated efficiently.

We acknowledge the support of ANPCyT, Argentina; YerPhI, Armenia; ARC, Australia; BMWFW and FWF, Austria; ANAS, Azerbaijan; SSTC, Belarus; CNPq and FAPESP, Brazil; NSERC, NRC and CFI, Canada; CERN; CONICYT, Chile; CAS, MOST and NSFC, China; COLCIENCIAS, Colombia; MSMT CR, MPO CR and VSC CR, Czech Republic; DNRF and DNSRC, Denmark; IN2P3-CNRS, CEA-DRF/IRFU, France; SRNSFG, Georgia; BMBF, HGF, and MPG, Germany; GSRT, Greece; RGC, Hong Kong SAR, China; ISF and Benoziyo Center, Israel; INFN, Italy; MEXT and JSPS, Japan; CNRST, Morocco; NWO, Netherlands; RCN, Norway; MNiSW and NCN, Poland; FCT, Portugal; MNE/IFA, Romania; MES of Russia and NRC KI, Russian Federation; JINR; MESTD, Serbia; MSSR, Slovakia; ARRS and MIZ\v{S}, Slovenia; DST/NRF, South Africa; MINECO, Spain; SRC and Wallenberg Foundation, Sweden; SERI, SNSF and Cantons of Bern and Geneva, Switzerland; MOST, Taiwan; TAEK, Turkey; STFC, United Kingdom; DOE and NSF, United States of America. In addition, individual groups and members have received support from BCKDF, CANARIE, CRC and Compute Canada, Canada; COST, ERC, ERDF, Horizon 2020, and Marie Sk{\l}odowska-Curie Actions, European Union; Investissements d' Avenir Labex and Idex, ANR, France; DFG and AvH Foundation, Germany; Herakleitos, Thales and Aristeia programmes co-financed by EU-ESF and the Greek NSRF, Greece; BSF-NSF and GIF, Israel; CERCA Programme Generalitat de Catalunya, Spain; The Royal Society and Leverhulme Trust, United Kingdom. 

The crucial computing support from all WLCG partners is acknowledged gratefully, in particular from CERN, the ATLAS Tier-1 facilities at TRIUMF (Canada), NDGF (Denmark, Norway, Sweden), CC-IN2P3 (France), KIT/GridKA (Germany), INFN-CNAF (Italy), NL-T1 (Netherlands), PIC (Spain), ASGC (Taiwan), RAL (UK) and BNL (USA), the Tier-2 facilities worldwide and large non-WLCG resource providers. Major contributors of computing resources are listed in Ref.~\cite{ATL-GEN-PUB-2016-002}.

\clearpage


\printbibliography

\clearpage 

 
\begin{flushleft}
{\Large The ATLAS Collaboration}

\bigskip

M.~Aaboud$^\textrm{\scriptsize 34d}$,    
G.~Aad$^\textrm{\scriptsize 99}$,    
B.~Abbott$^\textrm{\scriptsize 125}$,    
D.C.~Abbott$^\textrm{\scriptsize 100}$,    
O.~Abdinov$^\textrm{\scriptsize 13,*}$,    
B.~Abeloos$^\textrm{\scriptsize 129}$,    
D.K.~Abhayasinghe$^\textrm{\scriptsize 91}$,    
S.H.~Abidi$^\textrm{\scriptsize 164}$,    
O.S.~AbouZeid$^\textrm{\scriptsize 39}$,    
N.L.~Abraham$^\textrm{\scriptsize 153}$
% ...
\bigskip
\\

% ...
$^{13}$Institute of Physics, Azerbaijan Academy of Sciences, Baku; Azerbaijan.\\
% ...
$^{34}$$^{(a)}$Facult\'e des Sciences Ain Chock, R\'eseau Universitaire de Physique des Hautes Energies - Universit\'e Hassan II, Casablanca;$^{(b)}$Centre National de l'Energie des Sciences Techniques Nucleaires (CNESTEN), Rabat;$^{(c)}$Facult\'e des Sciences Semlalia, Universit\'e Cadi Ayyad, LPHEA-Marrakech;$^{(d)}$Facult\'e des Sciences, Universit\'e Mohamed Premier and LPTPM, Oujda;$^{(e)}$Facult\'e des sciences, Universit\'e Mohammed V, Rabat; Morocco.\\
% ...
$^{39}$Niels Bohr Institute, University of Copenhagen, Copenhagen; Denmark.\\
% ...
$^{91}$Department of Physics, Royal Holloway University of London, Egham; United Kingdom.\\
% ...
$^{99}$CPPM, Aix-Marseille Universit\'e, CNRS/IN2P3, Marseille; France.\\
$^{100}$Department of Physics, University of Massachusetts, Amherst MA; United States of America.\\
% ...
$^{125}$Homer L. Dodge Department of Physics and Astronomy, University of Oklahoma, Norman OK; United States of America.\\
% ...
$^{129}$LAL, Universit\'e Paris-Sud, CNRS/IN2P3, Universit\'e Paris-Saclay, Orsay; France.\\
% ...
$^{153}$Department of Physics and Astronomy, University of Sussex, Brighton; United Kingdom.\\
% ...
$^{164}$Department of Physics, University of Toronto, Toronto ON; Canada.\\
% ...
$^{a}$ Also at Borough of Manhattan Community College, City University of New York, NY; United States of America.\\
$^{b}$ Also at California State University, East Bay; United States of America.\\
$^{c}$ Also at Centre for High Performance Computing, CSIR Campus, Rosebank, Cape Town; South Africa.\\
$^{d}$ Also at CERN, Geneva; Switzerland.\\
$^{e}$ Also at CPPM, Aix-Marseille Universit\'e, CNRS/IN2P3, Marseille; France.\\
% ...
$^{*}$ Deceased

\end{flushleft}


\end{document}